\documentclass[english, 12pt]{article}

\RequirePackage[OT1]{fontenc}
\RequirePackage{amsthm,amsmath, amssymb}
\RequirePackage[round, longnamesfirst]{natbib}
\RequirePackage[colorlinks, linkcolor=blue, anchorcolor=blue, citecolor=blue,
filecolor=blue, urlcolor=blue]{hyperref}
\usepackage{bbm}
\usepackage{subfigure}
\usepackage{graphicx}
\usepackage[english]{babel}
\usepackage{setspace}
\usepackage[margin=1in]{geometry}

\usepackage{amsthm,amsfonts,amsmath,amssymb,bm}
\usepackage{lineno,hyperref}
\theoremstyle{definition}
\newtheorem{remark}{Remark}

\title{A defective cure rate quantile regression model for male breast cancer data}

\author{Agatha Rodrigues\thanks{Corresponding author. Email: agatha.rodrigues@ufes.com},  Patrick Borges and Bruno Santos \\ 
   {\small Department of Statistics, Federal University of Esp\'irito Santo, Vit\'oria - ES, Brazil}
}
\date{}

\begin{document}
\maketitle

%\begin{frontmatter}

\begin{abstract} 
In this article, we particularly address the problem of assessing the impact of clinical stage and age on the specific survival times of men with breast cancer when cure is a possibility, where there is also the interest of explaining this impact on different quantiles of the survival times. To this end, we developed a quantile regression model for survival data in the presence of long-term survivors based on the generalized distribution of Gompertz in a defective version, which is conveniently reparametrized in terms of the $q$-th quantile and then linked to covariates via a logarithm link function. This proposal allows us to obtain how each variable affects the survival times in different quantiles. In addition, we are able to study the effects of covariates on the cure rate as well. We consider Markov Chain Monte Carlo (MCMC) methods to develop a Bayesian analysis in the proposed model and we evaluate its performance through a Monte Carlo simulation study. Finally, we illustrate the advantages of our model in a data set about male breast cancer from Brazil.
\end{abstract}


%\end{frontmatter}
%\date{}

%-----------------------------------------------------------------------------------------------------------------------------------------------------
%-----------------------------------------------------------------------------------------------------------------------------------------------------
\section{Introduction}
\label{sec1}

Breast  cancer  is  the  second  most  frequent  cancer  type  that  occurs  in  women,  after  non-melanoma  skin neoplasms, but there is still some misunderstanding on its  incidence among men as well. According to the Centers for Disease Control and Prevention (CDC - \url{www.cdc.gov}), about 1 out of every 100 breast cancers diagnosed is found in a man, however, the occurrence of this neoplasm tends to increase due to the poor quality of life and the difficulty in making an early diagnosis. In a survey of scientific articles on breast cancer in men, \cite{HCS09} pointed out that the incidence has increased significantly from 0.86 to 1.06 per 100,000 men over the past few decades; the highest rates occurring in North America and Europe and the lowest rates in Asia. Additionally, it was observed that men with breast cancer have the worst overall survival rates in relation to women, but this is probably due to their older age at the time of diagnosis, which corresponds to the most advanced stage of presentation of the disease, as well as higher death rates due to disease comorbidity. %Agatha: Acho que essa frase não precisa. Cancer is an important public health concern around the world. 

In view of this problem, it is very desirable to establish which risk factors impact the survival of men with breast cancer, in order to understand how the disease works in this population. This information would allow physicians to prevent the overall progressive burden of this disease, with measures of control and preventive interventions in this context.

In population-based cancer survival studies, survival models that take into account a cure fraction, known as cure rate models or long-term survival models, are constantly used to simultaneously explain immune and susceptible patients to the cause of failure event of interest under study. The most common approach to the cure rate model is the standard mixture model approached by \cite{B49} and extended by \cite{berkson} and later extensively studied by \cite{F77}, \cite{F82}, \cite{F82}, \cite{MZ96}, \cite{BC04}, \cite{schnell15} and \cite{YT12}, among others. Although the mixture cure rate model is the most widely used in the literature, other alternative models of population modeling with cured individuals have been studied, such as: (i) models based on the latent competitive risk structure \citep{CO84}; (ii) models based on defective distributions, defined as distributions that are not normalized to one for some values of its parameters, a concept introduced by \cite{BDM09}. 

Based on item (i), for example, \cite{YT96} and \cite{CIS99} in cancer recurrence scenarios, assume that a latent biological process of propagation of latent clonogenic tumor cells (latent competitive risks) produces the observed failure (relapse), resulting in the model known in the literature as bounded cumulative hazard model. A unified approach of the standard mixture and bounded cumulative hazard models was proposed by \cite{YI05}. \cite{RCBC11} developed a more flexible cure rate survival model which includes a destructive process of the initial risk factors in a competitive scenario and is thus based on the biological mechanism of the occurrence of the event of interest. Subsequently, \cite{RCCB12} proposed a new Bayesian flexible cure rate survival model which generalizes the stochastic model of \cite{KRY93} and has much in common with the destructive cure rate model formulated by \cite{RCBC11}. \cite{BRB12} introduced a new cure rate survival model which extends the model of \cite{RCBC11} by incorporating a structure of dependence between the initiated cells. Recently, \cite{BRB16} proposed a cure rate survival hybrid model for accommodating characteristics of unobservable stages (initiation, promotion and progression) of carcinogenesis from survival data in the presence of latent competing risks. 

Regarding the item (ii), defective distributions have the advantage of allowing a cure rate without adding any extra parameters in the modeling, unlike above-mentioned cure rate models \citep{SCRLTR19}. In addition, there is no need to assume the existence of a cured fraction before modeling, the parameter estimates will tell whether or not there is a proportion of cured individuals. The two main distributions used for this purpose are the inverse Gaussian and Gompertz. Proposals dealing with the defective distributions have appeared in the literature, to name a few we refer to \cite{CS92}, \cite{GCRSP98}, \cite{RNTL16}, \cite{{B17}}, \cite{MA18}, \cite{SCRLTR19}  and \cite{CRTL19}.

All of the aforementioned survival models with a cure fraction are essentially models that focus on estimating the hazard ratio, scale, and cure fraction parameters but fail to address the quantile function estimate for any patient with arbitrarily chosen covariate values \citep{GCPS16}. For example, patients and doctors are often interested in this quantile function, because it allows doctors to make a quick assessment and prediction of a patient's survival time, rather than just the estimated probability of cure, that is, patients and doctors want to know how the regression coefficients of a given covariate change for different quantiles of survival time. Thus, it is of interest to investigate quantile regression models with a cure fraction.

Quantile regression, introduced by \cite{KB78}, is particularly useful when the rate of change in the conditional quantiles, expressed by the regression coefficients, depends on the quantile. One of its main advantages concerns its flexibility for modeling data with heteroscedasticity. Even though quantile regression for survival data have been extensively studied in the literature (e.g. \cite{YJW95}; \cite{ HKP02}; \cite{ P03}; \cite{YZL08};  \cite{PH08}; \cite{NJ13}; \cite{XXS18}), there are not many references dealing with quantile regression in a cure rate setup, known as cure rate quantile regression models; \cite{WY13} and \cite{GCPS16} could be mentioned as some of the examples. These proposals have included a non-parametric component either in the functional form of the regression equation or the distribution of the lifetime of the model, or both. However, to the best of our knowledge, no one has so far examined a particular fully parametric approach to cure rate quantile regression based on defective distributions, which, as mentioned earlier, have the competitive advantage of allowing a cure rate without adding any extra parameters in the modeling process. Therefore, the main objective of this paper is to consider a quantile regression model in which the lifetime variable is a defective distribution. Motivated by the work of \cite{B17}, we used as defective distribution the generalized Gompertz introduced by \cite{GHO13}, which is conveniently reparametrized in terms of the q-th quantile and then linked to covariates by means of a logarithm link function, which allow us to estimate the quantiles directly. Moreover, we explore the use of Markov chain Monte Carlo (MCMC) methods to develop a Bayesian analysis in the proposed model. 

An important detail of our approach is that we use a parametric quantile regression (PQR) model, similarly to what \cite{NJ13} considered for their generalized gamma proposal, though in a classic framework. This concept was thoroughly discussed by \cite{wg08}, where the author argued that the idea to model the quantile function of a probability distribution, instead of another location parameter, for instance, adds much more flexibility to one's modelling strategy. Specifically in comparison to the usual quantile regression defined by \cite{KB78}, this technique presents a great advantage, as we are able to avoid a big concern about quantile regression models, which is the issue with crossing quantiles. Because we are considering the defective distribution generalized Gompertz to model these quantiles and correspondingly its cure fraction, we coin this model as a defective cure rate quantile regression model. We believe this proposal brings more adaptability to current modeling efforts of survival times.

We organize the rest of the paper as follows: the model formulation is described in Section \ref{sec2}. Parameter inference under a Bayesian perspective is discussed in Section \ref{sec3}. In Section \ref{sec4}, we carry out a Monte Carlo simulation study to demonstrate the performance of the proposed estimation method. An application to men with breast carcinoma is discussed in Section \ref{sec5}. Furthermore, some conclusions are mentioned in Section \ref{sec6}.

%-----------------------------------------------------------------------------------------------------------------------------------------------------
%-----------------------------------------------------------------------------------------------------------------------------------------------------
\section{Generalized Gompertz Cure Rate Quantile Regression}
\label{sec2}

The generalized Gompertz (GG) distribution introduced by \cite{GHO13} considers that lifetime $T$ conditional to parameters $\lambda$, $\alpha$ and $\theta$, has density and survival functions, respectively, given by
\begin{equation}
\label{eq-1}
f(t\mid \lambda,\alpha,\theta)=\lambda\theta\exp\left\{\alpha t-\frac{\lambda}{\alpha}\left(\exp\{\alpha t\}-1\right)\right\}\left(1-\exp\left\{-\frac{\lambda}{\alpha}\left(\exp\{\alpha t\}-1\right)\right\}\right)^{\theta-1}, \quad t > 0 \end{equation}
and
\begin{equation}
\label{eq-2}
S(t\mid \lambda,\alpha,\theta)=1-\left(1-\exp\left\{-\frac{\lambda}{\alpha}\left(\exp\{\alpha t\}-1\right)\right\}\right)^\theta, 
\end{equation}
where $\lambda>0$ and $\alpha>0$ are scale parameters and $\theta>0$ is a shape parameter. The GG distribution includes the following distributions as special cases: (i) generalized exponential distribution \cite{GK99} when $\alpha$ tends to zero; (ii) Gompertz distribution when $\theta = 1$; and (iii) exponential distribution when $\alpha$ tends to zero and $\theta = 1$. 

The quantile and hazard functions of the GG distribution are given by
\begin{equation*}
\label{eq-3}
\mu_q=\mu(q\mid\lambda,\alpha,\theta)=\frac{1}{\alpha} \log\left(1 - \frac{\alpha}{\lambda} \log\left(1 - q^{1/\theta} \right) \right), \quad 0<q<1.
\end{equation*}
and
\begin{equation*}
\label{eq-4}
h(t\mid\lambda,\alpha,\theta)=\frac{\lambda\theta\exp\{\alpha t-\frac{\lambda}{\alpha}\left(\exp\{\alpha t\}-1\right)\}\left(1-\exp\{-\frac{\lambda}{\alpha}\left(\exp\{\alpha t\}-1\right)\}\right)^{\theta-1}}{1-\left(1-\exp\{-\frac{\lambda}{\alpha}\left(\exp\{\alpha t\}-1\right)\}\right)^\theta},
\end{equation*}
respectively. According to \cite{GHO13}, the shape of hazard function of the GG distribution is: (i) increasing if $\alpha>0$  and $\theta=1$; (ii) constant if $\alpha=0$ and $\theta=1$; (iii) increasing when $\theta>1$; (iv) decreasing if $\alpha=0$ and  $\theta<1$; and (v) bathtub if $\alpha>0$ and $\theta<1$. 

The GG distribution becomes a defective distribution, characterized by having a probability density function that integrates to values less than 1 \citep{BDM09}, if $\alpha<0$; thus being suitable for modeling survival data with long-term survivors. The corresponding cure fraction is:
\begin{equation}
\label{eq-5}
\lim_{t\rightarrow\infty}S(t\mid\lambda,\alpha,\theta)=1-\left(1-\exp\Bigg\{\frac{\lambda}{\alpha}\Bigg\}\right)^\theta=p_{0}(\lambda,\alpha,\theta)\in[0,1].
\end{equation}

\begin{remark}
\label{rmk1}
As highlighted in \citet{SCRLTR19}, unlike most cure rate models, defective distributions have the advantage of allowing a cure rate without adding any extra parameters in the modeling. In addition, there is no need to assume the existence of a cured fraction before modeling, the parameter estimates will tell whether or not there is a proportion of cured individuals. Nevertheless, as defective distributions do not use an additional parameter in modeling the cure fraction, their parameters will have "overloaded functions" in order to explain at the same time the characteristics of susceptible and immune individuals. It can cause concern if one of the study's objectives is to obtain some descriptive measures (for example, quantiles) of susceptible individuals, once the measures obtained in this context are not well defined, as the distribution is  defective. To circumvent this restriction, we will rewrite the defective distribution as a standard mixture model \citep {B49, berkson}, i.e., we will separate the susceptible and immune populations without including an additional parameter in the modeling; thus, in addition to ensuring the advantage of parsimony, we will also be able to describe the measures of interest to the susceptible population, which in our article are the quantiles.
\end{remark}

To consider a quantile regression model, due to the Remark \ref{rmk1}, first we write the model in \eqref{eq-2} as a standard mixture model \citep {B49, berkson}. In this case, we define an indicator variable $Z$, which takes the value 0 if the subject is immune and 1 if the subject is susceptible. Let $\mathbb{P}[Z=0]=p_{0}(\lambda,\alpha,\theta)$ and $\mathbb{P}[Z=1]=1-p_{0}(\lambda,\alpha,\theta)$, where $p_{0}(\lambda,\alpha,\theta)$ is given by \eqref{eq-5}. Then, the survival function in \eqref{eq-2} is given by
\begin{eqnarray}
\label{eq-new}
S(t\mid\lambda,\alpha,\theta)= p_{0}(\lambda,\alpha,\theta)+\big[1-p_{0}(\lambda,\alpha,\theta)\big]S_{1}(t\mid\lambda,\alpha,\theta)\mbox{,}
\end{eqnarray}
where $S_1(\cdot)$ denotes the survival function of the susceptibles. From  \eqref{eq-new}, we can get an expression for $S_{1}$ as
\begin{equation}
\label{eq-s1}
S_{1}(t\mid\lambda,\alpha,\theta)=\frac{S(t\mid\lambda,\alpha,\theta)-p_{0}(\lambda,\alpha,\theta)}{1-p_{0}(\lambda,\alpha,\theta)}.
\end{equation}

By considering the relation in \eqref{eq-s1}, we note that $S_{1}$ is a proper survival function with a quantile function given by
\begin{equation}
\label{eq-tq1}
\mu^{1}_{q}=\mu^{1}(q\mid\lambda,\alpha,\theta)=\frac{\log\left(\lambda-\alpha\log\left(1-q^{\frac{1}{\theta}}\left(1-\exp\{\frac{\lambda}{\alpha}\}\right)\right)\right)-\log(\lambda)}{\alpha},\quad 0<q<1.
\end{equation}

Next, we will reparameterize the density function in \eqref{eq-1} in terms of q-th quantile $\mu^{1}_{q}$, such that $\theta$ can be written as
\begin{equation}
\label{eq-theta}
\theta=-\frac{\log(q)}{\log\left(1-\exp\{\frac{\lambda}{\alpha}\}\right)-\log\left(1-\exp\left\{-\frac{\lambda}{\alpha}\left(e^{\alpha \mu_{q}^{1}}-1\right)\right\}\right)},
\end{equation}
where this result is obtained by isolating $\theta$ in Eq. \eqref{eq-tq1}. Then if we replace \eqref{eq-theta}  in \eqref{eq-1}, \eqref{eq-2} and \eqref{eq-5},  we have the reparameterized density and survival functions and cure fraction, which are given by, respectively, 

\begin{eqnarray}
\label{density_repar}
f(t\mid\lambda,\alpha,\mu_{q}^{1}) &=& \lambda\left[-\frac{\log(q)}{\log\left(1-\exp\{\frac{\lambda}{\alpha}\}\right)-\log\left(1-\exp\left\{-\frac{\lambda}{\alpha}\left(e^{\alpha \mu_{q}^{1}}-1\right)\right\}\right)}\right] \nonumber \\ 
&& \times \exp\{\alpha t-\frac{\lambda}{\alpha}\left(\exp\{\alpha t\}-1\right)\}\nonumber\\
&&\times\left(1-\exp\big\{-\frac{\lambda}{\alpha}\left(\exp\{\alpha t\}-1\right)\big\}\right)^{-\frac{\log(q)}{\log\left(1-\exp\{\frac{\lambda}{\alpha}\}\right)-\log\left(1-\exp\left\{-\frac{\lambda}{\alpha}\left(e^{\alpha \mu_{q}^{1}}-1\right)\right\}\right)}-1}\mbox{,} 
\end{eqnarray}
\begin{equation}
\label{survival_repar}
S(t\mid\lambda,\alpha,\mu_{q}^{1})=1-\left(1-\exp\big\{-\frac{\lambda}{\alpha}\left(\exp\{\alpha t\}-1\right)\big\}\right)^{-\frac{\log(q)}{\log\left(1-\exp\{\frac{\lambda}{\alpha}\}\right)-\log\left(1-\exp\left\{-\frac{\lambda}{\alpha}\left(e^{\alpha \mu_{q}^{1}}-1\right)\right\}\right)}}
\end{equation}
and
\begin{equation}
\label{cure_repar}
p_{0}(\lambda,\alpha,\mu_{q}^{1})=1-\left(1-\exp\Bigg\{\frac{\lambda}{\alpha}\Bigg\}\right)^{-\frac{\log(q)}{\log\left(1-\exp\{\frac{\lambda}{\alpha}\}\right)-\log\left(1-\exp\left\{-\frac{\lambda}{\alpha}\left(e^{\alpha \mu_{q}^{1}}-1\right)\right\}\right)}}.
\end{equation}
Hereafter, we shall use the notation $T\sim dGG(\alpha,\theta,\mu_{q}^{1}, q)$ where $\mu_{q}^{1}>0$ is the quantile parameter, $\lambda>0$, $\alpha < 0$, and $q\in (0,1)$ is known.

Now, we build the defective generalized Gompertz quantile regression model, imposing that the quantile $\mu_{q}^{1}$ of $T$ satisfies the following functional relation:
\begin{equation*}
%\label{reg_mu}
\mu_{q}^{1}(\boldsymbol{\beta},\textbf{x})= \mu_{q}^{1} =\exp(\textbf{x}^{\top}\boldsymbol{\beta}) \nonumber,
\end{equation*}
in which $\textbf{x}^{\top}=(1,x_{1},\ldots,x_{p})$ is  the vectors of covariates and  $\boldsymbol{\beta}=(\beta_{0},\beta_1,\ldots,\beta_{p})^{\top}$ is the unknown vector of regression parameters to be estimated. Thus, $\boldsymbol{\vartheta}_q=(\boldsymbol{\beta}^{\top},\lambda,\alpha)^{\top}$ is the vector of the parameters.

%-----------------------------------------------------------------------------------------------------------------------------------------------------
%-----------------------------------------------------------------------------------------------------------------------------------------------------
\section{Inference}
\label{sec3}

Consider that the lifetime $T$ is possibly not observed, that is, it is constrained by a right censored failure time and let $C$ denote the censoring time. In a sample of size $n$, we then observe the $i$-th lifetime $t_{i}=\min(T_i,C_i)$, and $i$-th failure indicator $\delta_i=I(T_i\leq C_i)$, where $\delta_i=1$ if $T_i$ is observed and $\delta_i=0$ otherwise, for $i=1,\ldots,n$. 

We consider that $T_i$'s are independent random variables with density and survival functions  $S \left( t_i\mid \boldsymbol{\vartheta}_q,{\bf x}_i \right)$ and $f \left( t_i\mid \boldsymbol{\vartheta}_q,{\bf x}_i \right)$, obtained by replacing $\mu_q^1=\exp(\textbf{x}^{\top}_i\boldsymbol{\beta})$ in Equations \eqref{density_repar} and \eqref{survival_repar}, respectively, where $\textbf{x}^{\top}_i=(1,x_{i1},\ldots,x_{ip})$, for $i=1,\ldots,n$, and $\boldsymbol{\vartheta}_q=(\boldsymbol{\beta}^{\top},\lambda,\alpha)^{\top}$ is a vector of unknown parameters for a given quantile $q$. We assume that each censoring time $C_i$ is independent of lifetime $T_i$, for all $i=1,\ldots,n$, and we consider a noninformative censoring assumption, i.e., the censoring distribution does not involve the parameters of the distribution of $T$. 
Therefore,  the likelihood function of $\boldsymbol{\vartheta}_q$ can be written as
\begin{equation*}
\label{eq-9}
L(\boldsymbol{\vartheta}_q;{\bf D})\propto\prod_{i=1}^{n}\left[f \left( t_i\mid \boldsymbol{\vartheta}_q,{\bf x}_i \right)\right]^{\delta_i}\left[S \left( t_i\mid \boldsymbol{\vartheta}_q,{\bf x}_i \right)\right]^{1-\delta_i},
\end{equation*}
where ${\bf D} = \left(n,{\bf t},{\boldsymbol \delta},{\bm X}\right)$, with ${\bf t} = \left(t_1, \ldots, t_n \right)^\top$, ${\boldsymbol \delta} = \left(\delta_1, \ldots, \delta_n \right)^\top$, and $ {\bm X}=(\textbf{x}^{\top}_1,\ldots,\textbf{x}^{\top}_n) $ is a $n\times p$ matrix  containing the covariate information.

Under the Bayesian approach, we can specify the posterior distribution of $\boldsymbol{\vartheta}_q$ as
 \begin{eqnarray} 
 \label{posteriori}
 \pi(\boldsymbol{\vartheta}_q \mid {\bf D}) & ~~ \propto &  ~~ \pi(\boldsymbol{\vartheta}_q)  L(\boldsymbol{\vartheta}_q;{\bf D}),    
 \end{eqnarray}
 where  $\pi(\boldsymbol{\vartheta}_q)$ is the prior distribution of $\boldsymbol{\vartheta}_q$. 
 
The prior distributions for $\boldsymbol{\vartheta}_q$ can be defined in the following manner, assuming that they are prior independent. Given our parameterization of the survival function of susceptible individuals in \eqref{eq-s1}, we should assume that $\alpha < 0$. Therefore, for $\alpha$ we assume a truncated normal distribution in the interval $(-\infty,0)$ with mean $-0.1$ and variance $100$. For the other parameters, we assume a gamma distribution with mean $1$ and variance $100$ for $\lambda$ and a normal distribution for each $\beta$ with mean $0$ and variance $100$. This setup will be considered in both the simulation and the application in the following sections.

The posterior distribution in \eqref{posteriori} does not have a closed form and the parameters are estimated through simulated samples of the posterior distribution obtained by the Adaptive Metropolis algorithm with multivariate normal distribution as proposed by \cite{Haario}. This approach is  implemented in the statistical package \textit{LaplacesDemon} \citep{laplacesDemon}, which provides a friendly environment for Bayesian inference within the R program \citep{R1}.
 
As a result, a sample of size $M$ from the joint posterior distribution of $\boldsymbol{\vartheta}_q$ is obtained (eliminating burn-in and jump samples). The sample from the posterior can be expressed as $(\boldsymbol{\vartheta}_q^{(1)},\boldsymbol{\vartheta}_q^{(2)},\ldots,\boldsymbol{\vartheta}_q^{(M)})$.

The posterior mean of $\boldsymbol{\vartheta}_q$, for instance, can be approximated by
\begin{eqnarray*}
\widehat{\boldsymbol{\vartheta}}_q=\frac{1}{M}\sum_{m=1}^{M}{\boldsymbol{\vartheta}_q^{(m)}}, \label{est_par_bayes}
\end{eqnarray*}
and the posterior mean of the cure rate is approximated by
\begin{eqnarray*}
\widehat{p}_0=\frac{1}{M}\sum_{m=1}^{M}{1-\left(1-\exp\Bigg\{\frac{\lambda^{(m)}}{\alpha^{(m)}}\Bigg\}\right)^{-\frac{\log(q)}{\log\left(1-\exp\{\frac{\lambda^{(m)}}{\alpha^{(m)}}\}\right)-\log\left(1-\exp\left\{-\frac{\lambda^{(m)}}{\alpha^{(m)}}\left(e^{\alpha^{(m)} \mu_{q}^{{1}^{(m)}}}-1\right)\right\}\right)}}}. \label{est_p0_bayes}
\end{eqnarray*}

\section{A simulation study}
\label{sec4}

In this section, we evaluate the performance of the proposed model considering a simulation study. We only consider one predictor variable, as we believe adding more variables would not cause problems for the estimation algorithm, though this could be further investigated in the future.

For this simulation scenario, we consider the following values for the parameters: $\alpha=-0.25$, $\lambda=1$, $\beta_0=1.3$, $\beta_1=0.7$ and $q \in\{0.2,0.5,0.8\}$. Besides, six sample sizes are considered, they are: $n=50,~ 100, ~ 300, ~ 500, ~1.000, ~2000$. For each combination of parameter values and sample size, $B=1000$ datasets are generated by considering the following algorithm:
\begin{enumerate}
	\item[1.] Determine desired values for $\alpha$, $\lambda$, $q$ and $\boldsymbol{\beta}=(\beta_0,\beta_1)^{\top}$;
	\item[2.] Define the proportion of censored data, given by $pc_0$ and $pc_1$, for $x=0$ and $x=1$, respectively;
	\item[3.] For the $i$th subject, draw $x_i\sim$ Bernoulli($0.5$), and calculate $p_{0x_i}$, in which $p_{00}$ and $p_{01}$ are given by (\ref{cure_repar}) when $x_i=0$ and $x_i=1$, respectively; 
	\item[4.] Draw $u_i\sim \mbox{Uniform}(0, 1)$. If $u_i< p_{0x_i}$, set $w_i = \infty$; otherwise, generate $u_{1i}\sim U(0,1-p_{0x_i})$ and calculate
	\begin{equation}
	   w_i=\frac{1}{\alpha}\log\Big[1-\frac{\alpha}{\lambda}\log\big(1-u_{1i}^{(1/\theta_{x_i})}\big) \Big], ~ \mbox{with} ~ x_i=1 ~ \mbox{or} ~ x_i=0; \nonumber
	\end{equation}
	\item[5.] Draw $c_i\sim U(0,\tau_i)$, where $\tau_i$ is defined to have approximately $p_{cx_i}$ proportion of censoring data.
	\item[6.] Determine $t_i=\min(w_i,ci)$. If $t_i=w_i$, set $\delta_i=1$, otherwise $\delta_i=0$; 
	\item[7.] Repeat steps 3 to 6 for all $i=1,\ldots,n$.  The data set for the $i$th subject is $\{t_i,x_i, \delta_i\}, \ i=1, \ldots, n$.
\end{enumerate}

It is worth mention that, when $q=0.2$ the cure fractions are $p_{00}=0.32$ and $p_{01}=0.83$; when $q=0.5$ the cure fractions are $p_{00}=0.15$ and $p_{01}=0.54$ and, finally, when $q=0.8$ the cure fractions are $p_{00}=0.05$ and $p_{01}=0.22$.

In order to obtain posterior quantities, the first $10,000$ samples were discarded as burn-out samples. A jump of size $10$ was chosen so that the correlation between the simulated values was close to zero. Thus, we get $M = 1,000$ simulated values for each parameter. We consider the posterior means for each parameter as our point estimates.

\begin{figure}[!h] 
	\begin{center}			\includegraphics[scale = 0.85]{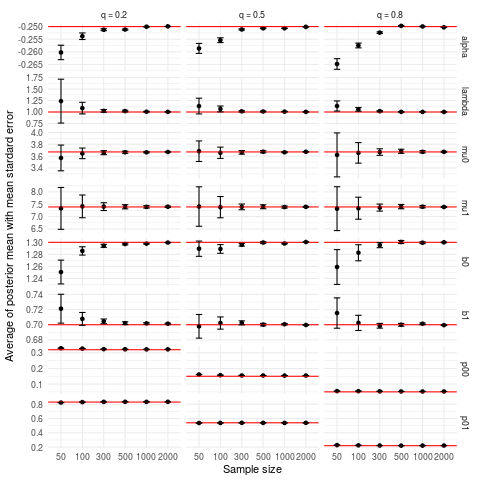}
		\caption{Posterior means for the different combinations of $q = \{0.2, 0.5, 0.8\}$, $n=50,~ 100, ~ 300, ~ 500, ~1000, ~2000$ and the different parameters represented by the dots. The red line represents the true value for each parameter. The bar around the dot denotes the mean squared error.}\label{bias_rmse}
	\end{center}
\end{figure}

In Figure~\ref{bias_rmse}, we can check that the bias decreases as the sample sizes increases, since the point estimates do approximate their respective true values. In particular to parameters $\alpha$ and $\beta_0$, for smaller sample sizes this bias is considerably large, given their respective mean squared error, though these values are minimal for sample size equal to 300. Moreover, we have that the mean squared error is relatively larger for small sample sizes, such as 50 and 100, but this number rapidly decreases as the sample sizes get larger as well. An important observation for this simulation study is that the cure fractions estimates present low bias values even for small sample sizes and for all $q$ considered. 

\begin{figure}[!h] 
	\begin{center}			\includegraphics[scale = 0.85]{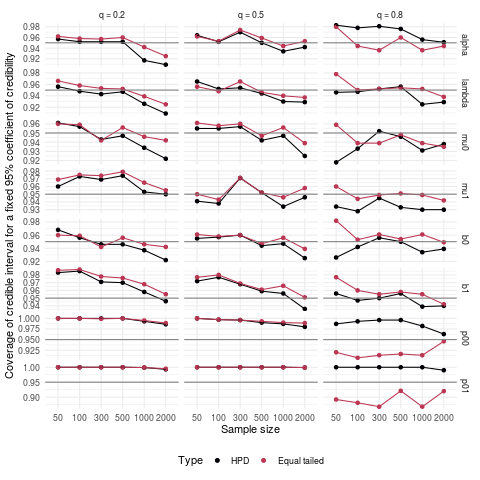}
		\caption{Coverage probability for the different combinations for $q = \{0.2, 0.5, 0.8\}$, $n=50,~ 100, ~ 300, ~ 500, ~1000, ~2000$ and the different parameters. The gray line represents the nominal value of 0.95.}\label{ci_sim}
	\end{center}
\end{figure}

In Figure~\ref{ci_sim} we compare the coverage probability of 95\% credible intervals for each parameter, considering HPD and equal tailed credible intervals. We can observe that the credible intervals for the cure fractions are mostly conservative, for $q=0.2$ and $q =0.5$. Additionally, for some parameters and for larger sample sizes, the coverage probability stays under the nominal value of 95\%. This is the case for example for $\alpha$ and $\lambda$ for $q=0.2$. Though this is not ideal, these estimates still do not vary very far from their nominal value.

%-----------------------------------------------------------------------------------------------------------------------------------------------------
%-----------------------------------------------------------------------------------------------------------------------------------------------------
\section{Application}
\label{sec5}

Breast cancer is the second most frequent cancer type that occurs in women, after non-melanoma skin neoplasms, but there is still some misunderstanding on its  incidence among men as well. According to the Centers for Disease Control and Prevention (\href{https://www.cdc.gov/cancer/breast/men}{www.cdc.gov}), about 1 out of every 100 breast cancers diagnosed is found in a man.

Breast cancer in women has been extensively studied, as well as its prognostic factors. The large number of studies is certainly comprehensible, as this type of cancer corresponds to almost 30\% of new cancer cases diagnosed annually in Brazilian women, being a major cause of death in the female population (according to Brazilian National Institute of Cancer - \href{https://www.inca.gov.br/}{www.inca.gov.br}). In this study, we want to study the impact of age and clinical stage in survival for men, in order to understand how the disease affects this population. 

For that, we consider a male breast cancer dataset from a  retrospective survey of 872 records of males diagnosed with breast in the state of S\~ao Paulo, Brazil, between 2000 and 2019, with follow-up conducted until February of 2020 and with at least two months of follow-up.

This dataset is provided by the Fundaç\~ao Oncocentro de S\~ao Paulo (FOSP), which is responsible for coordinating and monitoring the implementation of the Hospital Cancer Registry in the State of S\~ao Paulo (Brazil), in addition to systematizing and evaluating cancer care data available for the state. The FOSP is a public institution connected to the State Health Secretariat that monitors the evolution of the Oncology Care Network, assists the State Department of Health in the creation and application of prevention and health promotion programs, and monitors the evolution of cancer mortality in the state. The dataset considered here and the routines used to estimate the models are available in the following link: \url{https://github.com/brsantos/ggcrqr}.

Death due to breast cancer was defined as the event of interest. Those patients who did not die due to breast cancer during the follow-up period were characterized as right-censored observations. As cited before, the main goal is to assess the impact of clinical stage and age ($<$55; 55 - 65 and $>$65 years old) on the specific survival times. 

Of the 872 patients, 78\% did not die during the follow-up period, that is, 681 patients have a right-censored event time. A total of 268 (30.7\%) patients are under the age of 55, 275 (31.5\%) are between 55 and 65 years old and 329 (37.8\%) are older than 65 years old. Besides that, 21.3\% of the patients are classified as clinical stage I, 38\% as clinical stage II, 30.2\% as clinical stage III and 10.5\% as clinical stage IV. 

\begin{figure}[!h] \centering
	\begin{center}
		\begin{minipage}[b]{0.49\linewidth}
		\includegraphics[width=\linewidth]{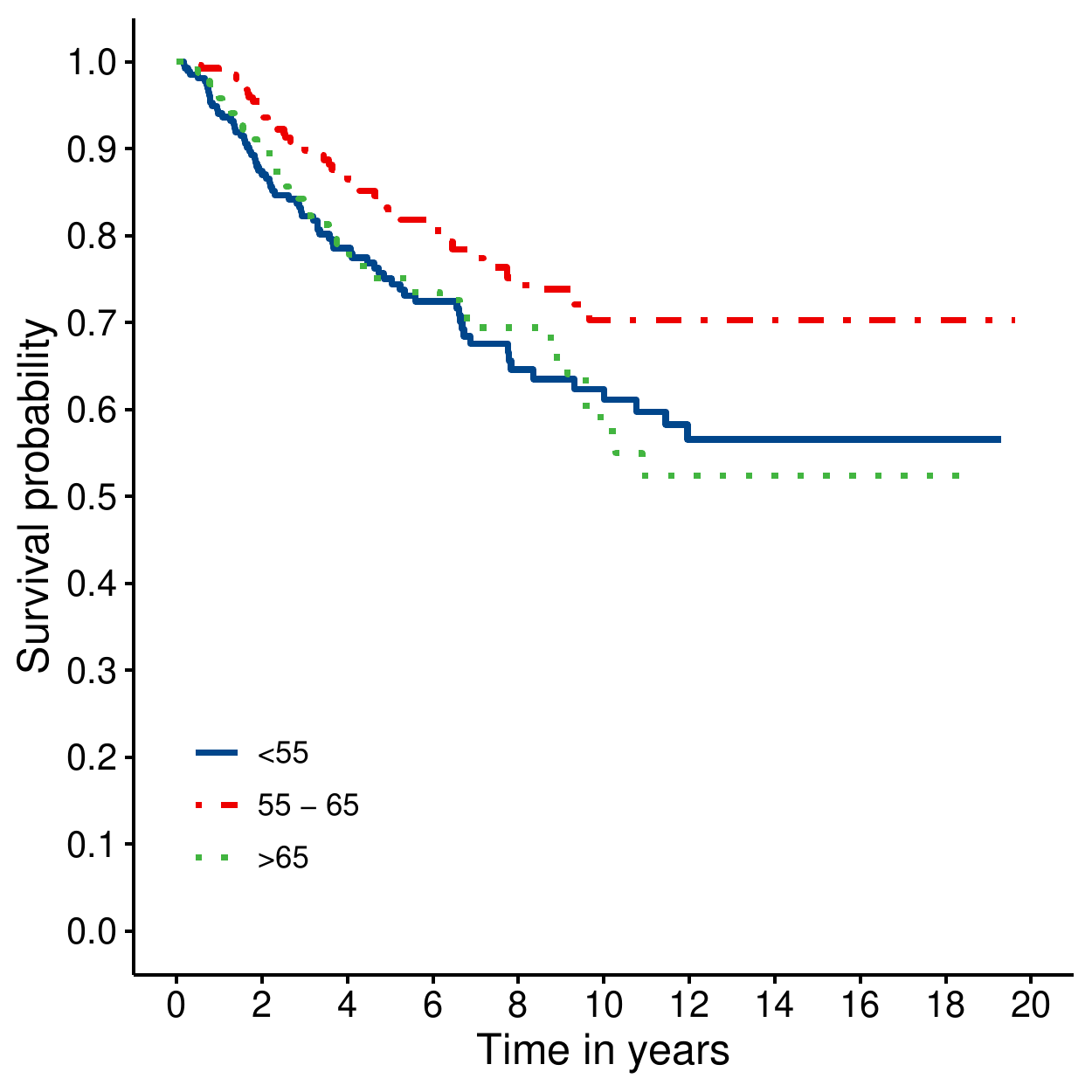}
%			\subcaption{ }
		\end{minipage}
		\begin{minipage}[b]{0.49\linewidth}
		\includegraphics[width=\linewidth]{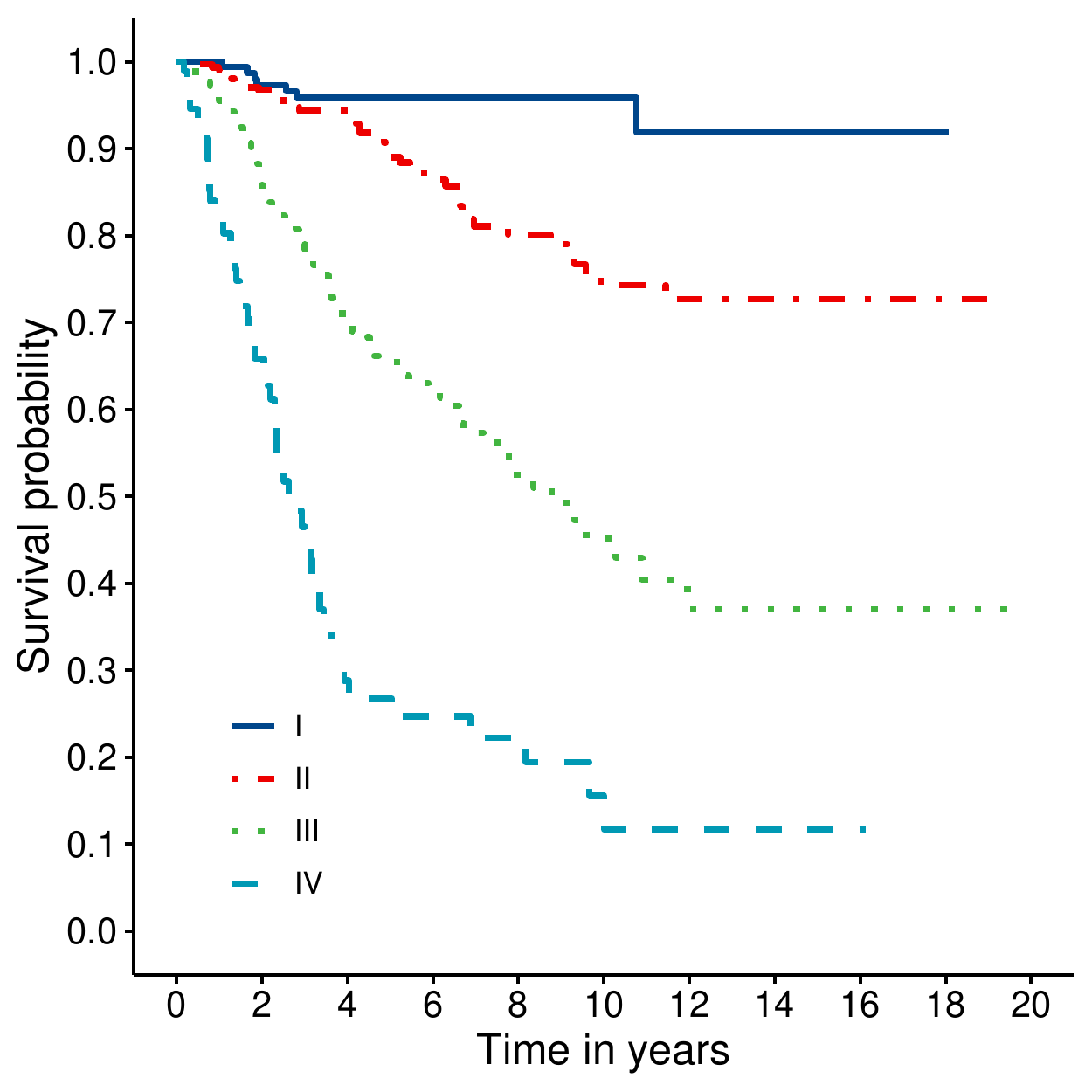}
        % \subcaption{ }
		\end{minipage}
		\caption{Kaplan-Meier estimates for the male breast cancer dataset grouped by age (in the left) and clinical stage (in the right).}\label{KM}
	\end{center}
\end{figure}

Figure \ref{KM} presents the Kaplan-Meier estimates for each explanatory variable. Of note, there is a strong evidence that a fraction of the population had been cured.
Among all of the variables considered in our study, those with clinical stage I had a better prognosis.

To evaluate the effects of clinical stage and age, the proposed model was fitted to the dataset. The quantile $\mu_q^1$ is given by:
\begin{equation}
\label{reg_mu_application}
\mu_{q}^{1}=\exp(\beta_0+\beta_1 \times \mbox{age}_{\geq 55 \& \leq 65} + \beta_2\times \mbox{age}_{> 65} +\beta_3 \times \mbox{stage}_{II} + \beta_4 \times \mbox{stage}_{III} +\beta_5 \times \mbox{stage}_{IV}),
\end{equation}
where we consider $q = \{0.05, 0.10, \ldots, 0.95\}$ and the coefficients $\beta_i$, $i=0,1,\ldots,5$ also vary with $q$, though we do not index them too to simplify the notation.

The adaptive Metropolis-Hastings algorithm was run, discarding the first $40{,}000$ iterations as burn-in samples and using a jump of size $70$ to avoid correlation problems, with a sample size of $n_p=1{,}000$. The convergence of the chain was evaluated by multiple runs of the algorithm from different starting values and was monitored through graphical analysis, where we were able to obtain good convergence results. 

\begin{figure}[!h] 
	\begin{center}
			\includegraphics[scale = 0.9]{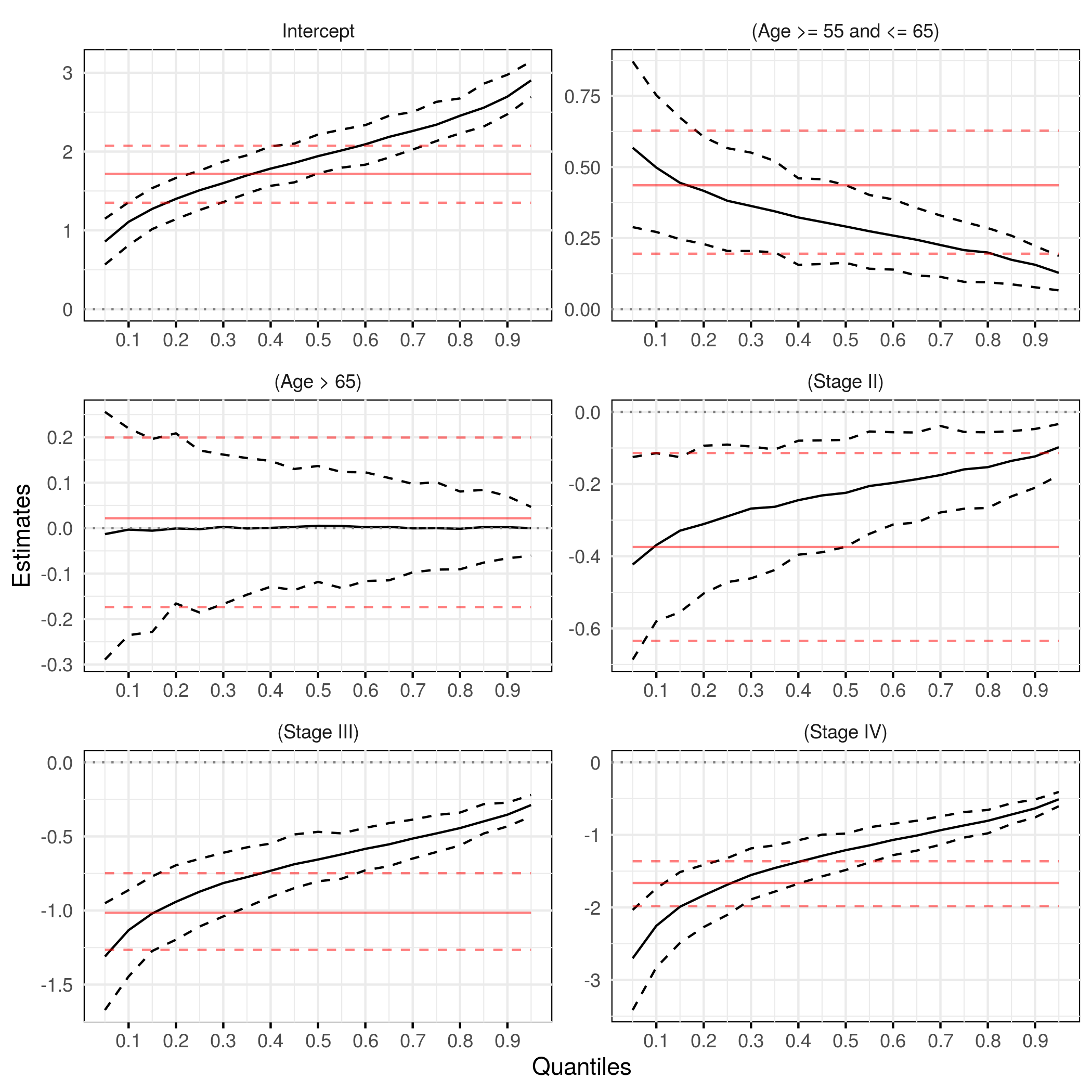}
		\caption{Posterior estimates for the regression parameter $\bf \beta$ given the quantiles, $q$. The black lines represent the estimates for the model in \eqref{reg_mu_application}. The red lines represent the estimates for the model in \eqref{reg_theta}. The solid lines denote the posterior mean, while the dashed are the respective HPD credible interval.} \label{coef_model}
	\end{center}
\end{figure}

Figure \ref{coef_model} presents the coefficients values by considering $q$ varying from $0.05$ to $0.95$ by $0.05$.  The solid line represents the posterior mean value and dotted lines represent the 95\% highest probability density (HPD) interval. The black lines represent the estimates for the proposed approach and the red lines the estimates of the model without the quantile transformation, i.e., considering the density and survival functions as in Equations \eqref{eq-1} and \eqref{eq-2}, respectively. Covariates are included in the latter model through the following equation:
\begin{equation}
\label{reg_theta}
\theta=\exp(\beta_0+\beta_1 \times \mbox{age}_{\geq 55 \& \leq 65} + \beta_2\times \mbox{age}_{> 65} +\beta_3 \times \mbox{stage}_{II} + \beta_4 \times \mbox{stage}_{III} +\beta_5 \times \mbox{stage}_{IV}).
\end{equation}

It is important to note how some coefficients vary along the different quantiles, which is not the case for all variables. For instance, for men with age over 65 years old, there is no evidence for quantiles of the survival times of susceptible individuals to be different from those with age under 55 years old, the reference category. This is indicated by the respective coefficient being close to zero for all values of $q$ considered. Also, for this variable there is no difference between the conclusions based on our proposed approach and the method based on $\theta$, given the similar point estimates for both models, except for the smaller credible interval presented by our proposal. This is slightly similar to the pattern of the parameter related to Stage II, though in this case, we find this parameter to be different than zero, given its credible interval. 

\begin{figure}[!h] 
	\begin{center}
		\includegraphics[scale = 0.9]{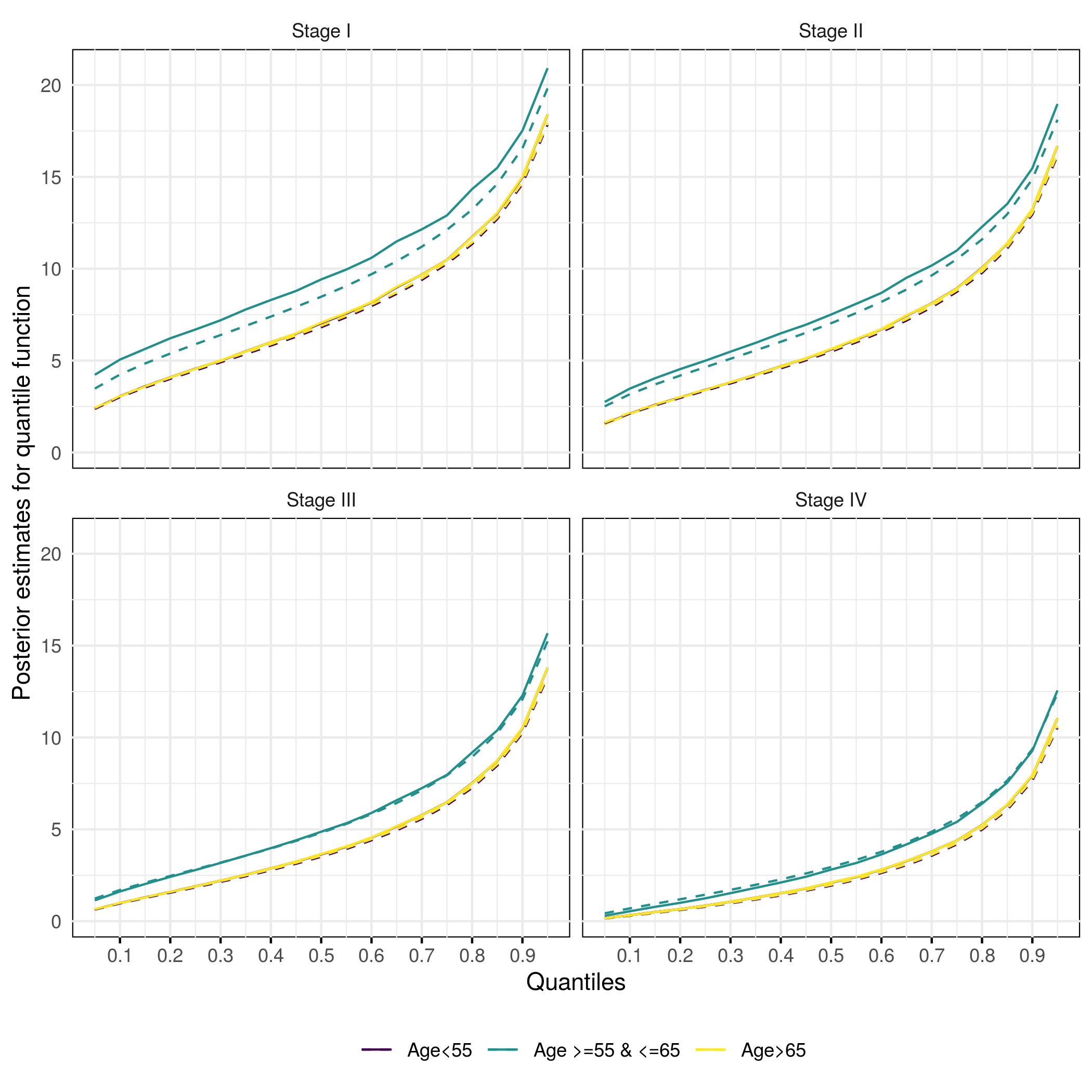}
		\caption{Posterior estimates for the quantile function for each combination of the predictor variables. The solid lines represent the estimates based on the quantile transformation, while the dashed lines represent the estimates based on without the quantile reparameterization.} \label{quantile_estimates}
	\end{center}
\end{figure}

A whole different conclusion can be said about the parameters associated with Stage III and Stage IV, with respect to Stage I defined as the reference category. The quantile-based model defined in \eqref{reg_mu_application} is able to estimate the effects for different quantiles and display contrasting differences along these quantiles, in comparison with the model for $\theta$. This is very substantial as we are able to conclude that though this effect is still negative for all considered $q$, the strength of this impact is smaller in absolute value for long term survival times, i.e., larger quantiles. Notice also that this is not always the case for the other parameters, as the credible intervals presented by the black lines and the red lines in Figure~\ref{coef_model} often intersect for all quantiles.

Figure~\ref{quantile_estimates} presents the quantile estimates for different values of $q$, obtained by the proposed model (solid lines) and without the quantile reparameterization (dashed lines), for the combination of clinical stage (I, II, III and IV) and age ($<$55; 55 - 65 and $>$65 years old). 

We can observe that for most combinations there is no difference between both parameterizations, which shows that both models are explaining the variation for survival times of the susceptible population similarly. This is an important consideration, given that with the quantile transformation we gather a more complete depiction of how each variable affects these quantiles locally, as we have discussed before. There are only small differences between estimates for men aged between 55 and 65 years old in Stage I and II, though these do not change the conclusions regarding the effects of each variable, corroborating with coefficients shown in Figure~\ref{coef_model}. For instance, while it seems that there is no line for men aged under 55 years old, in fact this is super imposed by the line for men aged over 65 years old, as it was discussed previously.

Another relevant feature on the results displayed on Figure~\ref{quantile_estimates} is the lack of quantile crossings, an issue often observed in the usual semiparametric quantile regression approach. This is certainly one of the advantages of the PQR method, which by definition avoids this issue. 

\begin{figure}[!h] 
	\begin{center}
		\includegraphics[scale = 0.6]{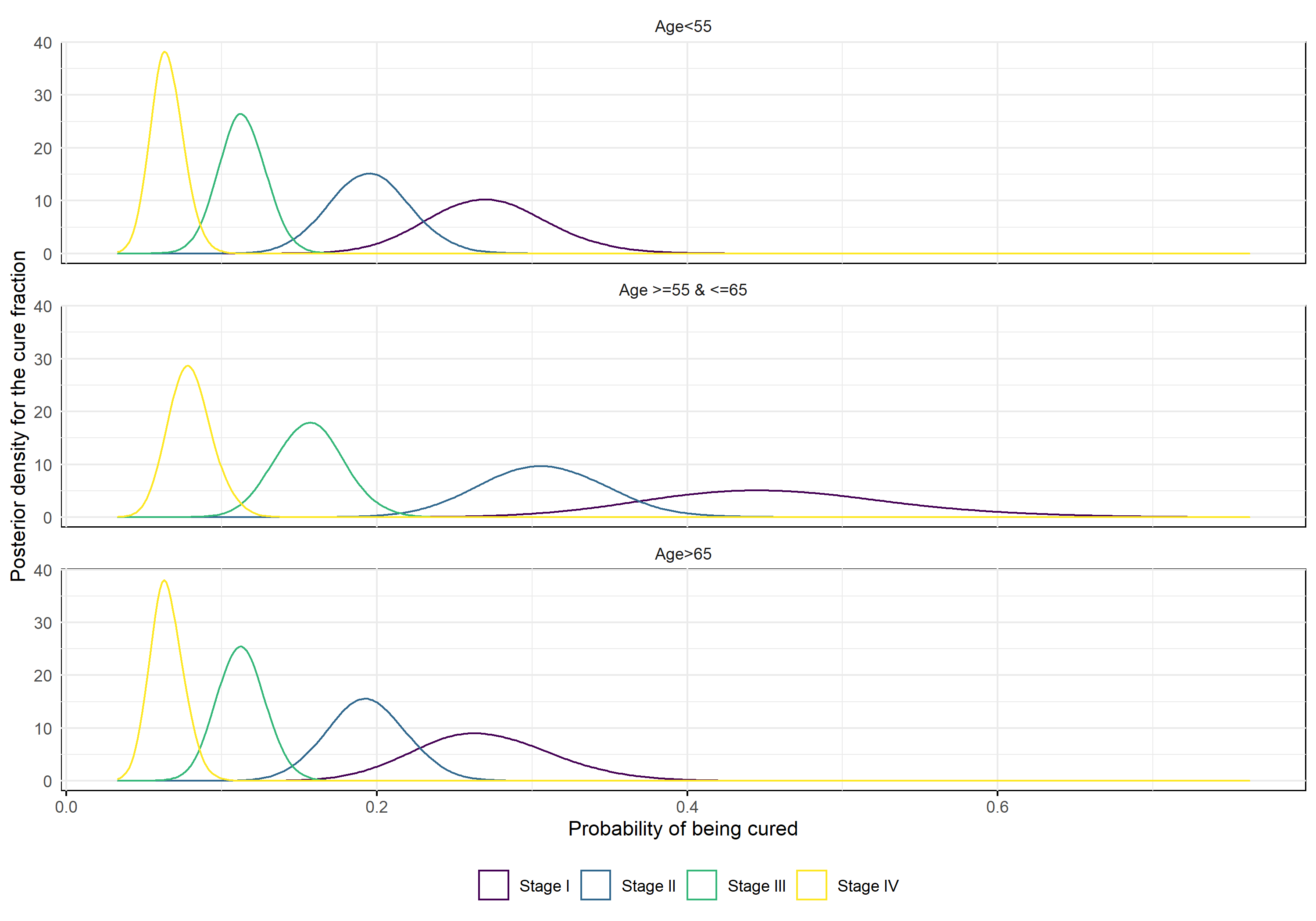}
		\caption{Posterior densities for the cure fraction parameter for each combination of covariates.} \label{prob_estimates}
	\end{center}
\end{figure}

Moreover, we can discuss the model estimates for the cure fraction with the formula given in \eqref{cure_repar}. Since we have the posterior draws for each parameter we can apply this function for each draw so we have a posterior sample from the cure fraction. This allows us to to check the posterior density for each combination of clinical stage and age, regarding its cure fraction as well. These densities are shown in Figure~\ref{prob_estimates} and the posterior mean as well as the 95\% credibility intervals for cure fraction are presented in Table  \ref{table_prob_cure}. % The analysis of this figure shows that differences between the distinct age groups is less pronounced than the different clinical stages.
 It is hard to point out the differences between older or younger men with regards to their cure probability, regardless of the disease clinical stage. In fact, if we combine the curves as in Figure~\ref{quantile_estimates} we see this more clearly, though we do not show it here for the sake of brevity. Overall for stages I to IV, men under older than 55 and younger than 65 years old have a higher probability of cure, in comparison with the other groups.
 
The ordering of the probability of cure for the different clinical stages is consistent with the common prognosis of this carcinoma, as people in Stage I have a higher probability of being cured, and men at Stage IV of the cancer diagnosis have a lower probability of being cured in comparison to other stages.  It is worth noting the high posterior variability in the distribution of the cure fraction for stage I aged in the interval $\geq$55 \& $\leq$65. A reason for this may be the extremely low occurrence of events for early clinical stages.

\begin{table}[]
    \centering
\begin{tabular}{ccccc}
	\hline
	Stage & Age & Posterior Mean & \multicolumn{2}{c}{95\% Credibility interval} \\  
	\hline
	Stage I  & Age$<$55  &        0.272  & 0.247 &  0.347 \\ 
    Stage I &  Age $\geq$55 \& $\leq$65 & 0.456 &  0.404  & 0.618 \\
    Stage I  &  Age$>$65   &       0.270 & 0.241 & 0.356 \\
    Stage II  & Age$<$55   &       0.195 & 0.179 & 0.245 \\
   Stage II  & Age $\geq$55 \& $\leq$65 & 0.307 & 0.281 & 0.384 \\
   Stage II & Age$>$65   &       0.193 &  0.177 & 0.241 \\
   Stage III & Age$<$55  &        0.113  & 0.104 & 0.142 \\
   Stage III & Age $\geq$55 \& $\leq$65 & 0.157 & 0.143 & 0.200 \\
  Stage III & Age$>$65  &        0.112  & 0.102 & 0.141 \\
  Stage IV  & Age$<$55     &     0.065 & 0.058 & 0.086 \\
   Stage IV & Age $\geq$55 \& $\leq$65 & 0.080 & 0.071 & 0.109 \\
   Stage IV &  Age$>$65     &     0.065 & 0.058 & 0.086 \\
   \hline
\end{tabular}
\caption{Posterior mean and 95\% posterior credibility interval for the cure fraction for each combination of covariates.}
\label{table_prob_cure}
\end{table}

%-----------------------------------------------------------------------------------------------------------------------------------------------------
%-----------------------------------------------------------------------------------------------------------------------------------------------------
%\clearpage

\section{Concluding Remarks}
\label{sec6}

While there is a lot of studies for female breast cancer, the data about this carcinoma on the male population is still very limited \citep{YIB16}. Here we provide more information on this disease in the Brazilian population of men, where we discuss the association of clinical stage and age on the survival times of men enduring this type of cancer. Our approach considered the use of a parametric quantile regression model, where we can observe how these variables affect the different quantiles of the survival times. By selecting a defective distribution in the form of the Generalized Gompertz distribution we are able to explain simultaneously both the quantiles of the survival times and the cure fraction, given the covariates, without adding new parameters to the probability distribution.

We have showed that our estimation approach based on MCMC draws effectively reach the correct values in a simulation study, while also presenting its main advantages in the application. Here we discussed how the quantile parameterization specifies a more complete picture of conditional distribution of survival times, as we can study how each variable influence the different quantiles of this distribution. This is accomplished without changing the possible conclusions based on a different parameterization, as we showed that the conditional quantiles obtained given another parameterization are similar to the ones based on the quantile parameterization, though one has a much better interpretation of the parameters in the latter. The parametric quantile regression approach also avoids the issue of quantile crossing, which is a very common problem discussed in the literature for these models. 

\section*{Acknowledgements}
The authors thank the Funda\c c\~ao Oncocentro de S\~ao Paulo for providing the men breast cancer dataset.

\end{document}